\documentstyle[prb,aps,multicol,epsfig]{revtex}
\begin{document}

\title{A brief note about V$_{2}$O$_{3}$. }
\author{S. Di Matteo$^{1,2}$}
\address {$^{1}${Laboratori Nazionali di Frascati INFN, via E. Fermi 40 I-00044
Frascati (Roma) Italy}\\
$^{2}${Dipartimento di Fisica, Universit\`a di Roma III, via della Vasca
  Navale 84, I-00146 Roma Italy}\\}
\date{\today}
\maketitle
\draft
\begin{abstract}
This note summarizes some recently published results, that are reported in cond-mat today. Its aim is twofold. First, I believe that it is worthwhile to clarify the theoretical interpretation of a series of x-ray scattering experimental results, whose implications are apparently not well-known in the recent literature. A comment about K edge linear dichroism experiments is also provided. 
In second place, I would like to add a personal opinion about the role of non-local correlations in the insulating ground-state of V$_{2}$O$_{3}$.
\end{abstract}

\vspace{1cm}

This note is written as a brief summary of the results published in [\onlinecite{dima1,dima2,dima3,dima4}], that are reported in cond-mat today. As such, I shall avoid any introduction, referring, eg, to [\onlinecite{dima2,dima3}].

{\it a) -- Theoretical information inferred from non-resonant x-ray magnetic scattering\cite{paolasini} and resonant x-ray scattering\cite{paolasini,paolasini1}.} 
Non-resonant magnetic scattering measures L/S.\cite{blume1} Combining the value\cite{paolasini} L/(2S)=$-$0.3 with the value L+2S=1.2 $\mu_B$ from neutron scattering,\cite{moon} we get S=0.85 $\mu_B$ and L=-0.5 $\mu_B$. Notice that non-resonant magnetic scattering does not create intermediate core holes and the previous values for L and S refer to the ground state. This implies that the atomic spin is more compatible with S=1 than S=1/2, as presently widely accepted.\cite{mila,dimatteo,tanaka} However, it also implies that the orbital angular momentum is not quenched, but has a sizeable magnitude, and therefore the orbital part of the wavefunction must be complex.

The results of RXS ab-initio simulations are described in [\onlinecite{dima2}]: there it is shown that the orbital ordering cannot be responsible for the signal, contrary to what previously believed.\cite{paolasini,mila}
The signal at, eg, the (111)$_m$ reflection is found to be due mainly to a E1-E2 magnetic component, with a smaller E2-E2 contribution. This implies again, and independently, that the orbital ground state is complex. In fact, when the spin-orbit interaction in $3d$ vanadium states was "switched off" in our numerical simulation, no signal was recovered for all reflections with $h+k+l$=odd.
To my knowledge, the only model available in the literature that takes correclty into account spin-orbit coupling of $3d$ states in V$^{3+}$-ion is the one by A. Tanaka.\cite{tanaka}

Finally, in [\onlinecite{dima2}] it was also shown that the absence of any signal in the $4p$ energy region is a definitive proof that the magnetic moment has no component out of the glide plane, thus solving an old controversy of polarised neutron results.\cite{moon,word}

\vspace{0.5cm}

{\it b) -- K edge linear dichroism.} In Refs. [\onlinecite{dima1,dima4}], we report our conclusions about possible non-reciprocal effects in the ground state of V$_2$O$_3$. The problem was raised by the K edge linear dichroism experiment described in Ref. [\onlinecite{goulon}], whose interpretation seemed to imply that the ground state of V$_2$O$_3$ in its AFI phase was magnetoelectric. This, in turn, would have implied a different orbital simmetry than what usually believed, as reported in Ref. [\onlinecite{dimatteo}].
However, no magnetoelectricity has been detected in the system.\cite{astrov}  
The conclusions of both our theoretical\cite{dima1} and experimental \cite{dima4} works are that the non-reciprocal effect is probably an artefact of the experimental setup and it is anyway not related to the AFI ground state.

\vspace{0.5cm}

{\it c) -- Considerations on the molecular wavefunctions.} A question: why is the magnetic moment totally lying on the glide plane ? From symmetry arguments related to the single V$^{3+}$-ion, there are no reasons for this fact (the atomic symmetry is ${\hat{C}}_1$). If, however, the measured spin did not belong to the V$^{3+}$-ion, but to the "vertical" molecule, its local symmetry would not admit an out-of-plane component (see Sec. 4 of [\onlinecite{dima3}]). This seems to imply that the fundamental unit of the system is the molecule rather than the ion.\cite{note}

Another argument pointing towards the vertical molecule comes from the energetics (see again Sec. 4 of [\onlinecite{dima3}] for a short description and Sec. V of [\onlinecite{dimatteo}] for more detailed calculations). With the available values of the hopping integrals,\cite{mattheiss} the entangled form of the vertical molecule gains about 90 meV/atom (!) from non-local correlations.

This form of energy gain, unfortunately, cannot be taken into account by LDA-based approaches, which, by definition, have a local exchange and correlation potential. My opinion, in this sense, is that, in order to obtain a better comprehension of V$_2$O$_3$ ground-state, one should look to a non-local approximation of DFT equations, more than insisting with local self-energies, even though as accurate as LDA+U, LDA+DMFT, etc.

\vspace{0.5cm}

{\it Warning} -- The V$_2$O$_3$ problem is of course much wider than what discussed here and in Refs. [\onlinecite{dima1,dima2,dima3,dima4,dimatteo}], where just some features of the insulating phases are described. The PM-PI transition is not dealt with, and, moreover, one should keep in mind that the x-rays spectroscopies [\onlinecite{dima4,paolasini,paolasini1,goulon}] are performed with 2.8$\%$ Cr-doped single crystals, in the PI-AFI region of the phase diagram, in order to avoid the desruptive first-order phase transition of the pure sample. It is always tacitly assumed that such a small Cr-doping does not alter the main characteristics of the electronic ground state of pure V$_2$O$_3$. However, the only experimental paper in which this point is analyzed in some detail, is, to my knowledge, Ref. [\onlinecite{yethiraj}].

\end{document}